\documentclass[useAMS,usenatbib]{mn2e}

\usepackage{amsmath,amssymb,amsfonts,latexsym} 
\usepackage[dvips]{graphicx}
\usepackage{longtable}
\usepackage{fixltx2e}
\usepackage{lscape}

\title{On the pulse--width statistics in radio pulsars.\\ 
II. Importance of the core profile components}

\author[Krzysztof Maciesiak \& Janusz Gil]{Krzysztof
Maciesiak$^{1}$\thanks{E-mail:jezyk@astro.ia.uz.zgora.pl}, Janusz
Gil$^{1}$\\
$^{1}$Kepler Institute of Astronomy, University of Zielona G\'{o}ra, Lubuska 2, 65-265
Zielona G\'{o}ra, Poland}
\begin{document}

\date{Accepted . Received ; in original form }

\pagerange{\pageref{firstpage}--\pageref{lastpage}} \pubyear{2011}

\maketitle

\label{firstpage}

\begin{abstract}
We performed a statistical analysis of half-power pulse--widths of the core components in average pulsar profiles. We confirmed an existence of the lower bound of the distribution of half-power pulse--width versus the pulsar period $W_{50}\sim2^{\circ}.45 \; P^{-0.5}$ found by Rankin (1990). Using our much larger database we found $W_{50}=(2^{\circ}.51\pm0^{\circ}.08) \; P^{-0.50\pm0.02}$ for 21 pulsars with double--pole interpulses for which measurement of the core component width was possible. On the other hand, all single--pole interpulse cases lie in the swarm of pulsars above the boundary line. Using the Monte Carlo simulations based on exact geometrical calculations we found that the Rankin's method of estimation of the inclination angle $\alpha\approx asin(2^{\circ}.45\; P^{-0.5}/W_{50})$ in pulsars with core components is quite good an approximation, except for very small angles $\alpha$ in almost aligned rotators.
\end{abstract}


\section{Introduction}
It is generally accepted that the elements of mean pulsar profiles can be divided into core and conal components (\citet{r83}, Rankin (1990, hereafter \citet{r90}), \citet{r93}). The core component usually occupies central parts of the profile and is often flanked by one or two pairs of conal components. The importance of the conal profile components was recently studied statistically by Maciesiak, Gil \& Ribeiro (2011; hereafter \citet{mgr11}). These authors used recently large data sets of pulsar periods, profile widths and inclination angles resulted from large pulsar surveys conducted in recent years. They also compiled the largest ever database of pulsars with interpulses (IP), divided into the double--pole (DP--IP) and single--pole (SP--IP) cases (see Table \ref{tab.1} in this paper). The IP pulsars constitute about 3 percent (2 and 1 percent for DP--IP and SP--IP cases, respectively) of the pulsar population. The DP--IP pulsars (almost orthogonal rotators) tend to be younger or middle aged, while SP--IP pulsars tend to be older (see Figure 6 in \citet{mgr11}). This suggest a secular alignment of the magnetic axis towards the spin axis over a timescale of about $10^7$ years.

The interpulse emission is important also for the statistics of the core profile components in pulsars. Rankin (1990; hereafter R90) measured the half-power (50\% of the maximum intensity) widths of core components in a number of pulsars (about 100), including those showing the interpulse emission. She found that the distribution of pulse--widths has a clear lower bound $2^{\circ}.45 \; P^{-0.5}$ populated by DP--IP cases, in which the inclination angle $\alpha$ must be close to $90^{\circ}$. The rest of pulsars including SP--IP cases lied in a swarm of data points above the lower bound. \citet{r90} described this distribution by a simple mathematical expression 
\begin{equation}
      W_{50} = 2^{\circ}.45 \; P^{-0.5} / sin \alpha,
\label{eq.1}
\end{equation}
indicating that the core component widths depend only upon the pulsar period $P$ and the inclination angle $\alpha$ between the rotation and magnetic axes of the neutron star and they are almost insensitive to the impact angle $\beta$ of the closest approach of the observer's line--of--sight to the magnetic axis (for description of pulsar geometry see Paper I or Figure \ref{figure.b1} in the on-line Appendix \ref{appendix.b}). The above equation represents a useful method of estimation of the inclination angle $\alpha$ in pulsars with core component in the mean profile. In this paper we attempt to revisit and verify these important and intriguing results, using a more abundant database that we compiled in \citet{mgr11}).

\section{Importance of the core pulsar emission}

\begin{table*}
 \centering
 \begin{minipage}{155mm}
  \caption{Table of 44 known pulsars with interpulses divided into double--pole (DP--IP) and single--pole (SP--IP) cases. Bibliographic marks for interpulse information are the following: T/K -- \citet{kgm04} using \citet{taylor93}, WJ -- \citet{wj08}, A -- \citet{damico98}, M02 -- \citet{morris02}, K03 -- \citet{kramer03}, H -- \citet{hobbs04}, L -- \citet{lorimer06}, R -- \citet{ribeiro08}, M01 -- \citet{manchester01}, J -- \citet{janssen09}, K09 -- \citet{keith09}, K10 -- \citet{keith10}, C -- \citet{camilo09}, M11 -- Maciesiak et al. (2011). \label{tab.1}}
  \begin{tabular}{|r|l|l|r|c|c|c|c|c|}
\hline
\multicolumn{1}{|c|}{} & \multicolumn{1}{|c|}{}  & \multicolumn{1}{|c|}{} & \multicolumn{1}{|c|}{}& \multicolumn{3}{|c|}{$W_{50}$ [deg]}  & \multicolumn{1}{|c|}{SP--IP} & \multicolumn{1}{|c|}{} \\ \cline{5-7}

\multicolumn{1}{|c|}{No.} & \multicolumn{1}{|c|}{Name J}  & \multicolumn{1}{|c|}{Name B} & \multicolumn{1}{|c|}{Period}& \multicolumn{1}{|c|}{}& \multicolumn{1}{|c|}{} &\multicolumn{1}{|c|}{}& \multicolumn{1}{|c|}{or}    & \multicolumn{1}{|c|}{Bibliography} \\

\multicolumn{1}{|c|}{} & \multicolumn{1}{|c|}{}  & \multicolumn{1}{|c|}{} & \multicolumn{1}{|c|}{[s]}& \multicolumn{1}{|c|}{$2^{\circ}.45\: P^{-0.5}$}& \multicolumn{1}{|c|}{R90\footnote{R90 lists 6 cases of DP--IP in her Table 3 that she used to find a formal fit, but we were not able to confirm the interpulse in B033-45 (Vela pulsar). \citet{r90} in Table 2 lists few more DP--IP cases for which precise measurements of $W_{50}$ were not possible, so they were not included in her fit. Two additional cases B$1055-52$ and B$1822-09$ were taken from Table 5 in R90. \citet{r90} listed only one case of SP--IP.}} & \multicolumn{1}{|c|}{this paper} & \multicolumn{1}{|c|}{DP--IP} & \multicolumn{1}{|c|}{} \\\hline

1. & J$0534+2200$ & B$0531+21$& 0.033  & 13.49 & $13.5\pm1$  &            & DP & T/K \\
2. & J$0627+0706$ &           & 0.476  & 3.55  &             &            & DP & K10 \\
3. & J$0826+2637$ & B$0823+26$& 0.531  & 3.36  &$3.38\pm0.1$ &$3.47\pm1$  & DP & T/K \\
4. & J$0834-4159$ &           & 0.121  & 7.04  &             &$7.0\pm1$   & DP & K03\\
5. & J$0842-4851$ & B$0840-48$& 0.644  & 3.05  &             &$3.1\pm0.5$ & DP & M11\\
6. & J$0905-5127$ &           & 0.346  & 4.17  &             &$5.0\pm1$   & DP & WJ \\
7. & J$0908-4913$ & B$0906-49$& 0.107  & 7.49  &$7.5\pm0.4$  &$7.5\pm1$   & DP & T/K \\
8. & J$1057-5226$ & B$1055-52$& 0.197  & 5.52  &$<6$         &$5.8\pm1$   & DP & T/K \\
9. & J$1126-6054$ & B$1124-60$& 0.203  & 5.44  &             &            & DP & WJ \\
10.& J$1244-6531$ &           & 1.547  & 1.97  &             &            & DP & K09\\
11.& J$1413-6307$ & B$1409-62$& 0.395  & 3.90  &             &$ 4.1\pm1$  & DP & M11\\
12.& J$1549-4848$ &           & 0.288  & 4.57  &             &$ 5.0\pm1$  & DP & A\\
13.& J$1611-5209$ & B$1607-52$& 0.182  & 5.74  &             &$ 5.1\pm1$  & DP & WJ \\
14.& J$1613-5234$ &           & 0.655  & 3.03  &             &            & DP & R -- M01 \\
15.& J$1627-4706$ &           & 0.141  & 6.52  &             &            & DP & R -- L \\
16.& J$1637-4553$ & B$1634-45$& 0.119  & 7.10  &             &            & DP & WJ \\
17.& J$1705-1906$ & B$1702-19$& 0.299  & 4.48  &$4.5\pm0.3$  &$ 5.5\pm1$  & DP & T/K \\
18.& J$1713-3844$ &           & 1.600  & 1.94  &             &$ 2.1\pm0.5$& DP & K03\\
19.& J$1722-3712$ & B$1719-37$& 0.236  & 5.04  &             &$ 6.0\pm1$  & DP & T/K \\
20.& J$1737-3555$ & B$1734-35$& 0.398  & 3.88  &             &$ 4.65\pm1$ & DP & M11\\
21.& J$1739-2903$ & B$1736-29$& 0.323  & 4.31  &             &$ 4.8\pm1$  & DP & T/K \\
22.& J$1825-0935$ & B$1822-09$& 0.769  & 2.79  &$2.8$        &$2.8\pm0.25$& DP & T/K \\
23.& J$1828-1101$ &           & 0.072  & 9.13  &             &            & DP & M02 \\
24.& J$1842+0358$ &           & 0.233  & 5.08  &             &$7.0\pm1$   & DP & R -- L\\
25.& J$1843-0702$ &           & 0.192  & 5.59  &             &$5.8\pm1$   & DP & H\\
26.& J$1849+0409$ &           & 0.761  & 2.81  &             &$3.0\pm0.6$ & DP & L\\
27.& J$1913+0832$ &           & 0.134  & 6.69  &             &            & DP & M02 \\
28.& J$1915+1410$ &           & 0.297  & 4.49  &             &            & DP & R -- L\\ 
29.& J$1932+1059$ & B$1929+10$& 0.227  & 5.14  & $5.15\pm0.1$&$5.3\pm0.2$ & DP & T/K \\
30.& J$2047+5029$ &           & 0.446  & 3.67  &             &$3.88\pm0.5$& DP & J\\ 
31.& J$2032+4127$ &           & 0.143  & 6.47  &             &            & DP & C \\
\hline
1. & J$0828-3417$ & B$0826-34$& 1.849  & 1.80  &$40$         &$ 70$     & SP & T/K \\
2. & J$0831-4406$ &           & 0.312  & 4.39  &             &          & SP & R -- K03 \\
3. & J$0953+0755$ & B$0950+08$& 0.253  & 4.87  &             &$ 30$     & SP & T/K \\
4. & J$1107-5907$ &           & 0.253  & 4.87  &             &$ 50$     & SP & R -- L \\
5. & J$1302-6350$ & B$1259-63$& 0.048  & 11.21 &             &$ 39$     & SP & T/K \\
6. & J$1424-6438$ &           & 1.024  & 2.42  &             &          & SP & R -- K03 \\
7. & J$1637-4450$ &           & 0.253  & 4.87  &             &$ 19$     & SP & R -- L \\
8. & J$1806-1920$ &           & 0.880  & 2.61  &             & $20$     & SP & M02 \\
9. & J$1808-1726$ &           & 0.241  & 4.99  &             &$ 97$     & SP & L\\
10.& J$1851+0418$ & B$1848+04$& 0.285  & 4.59  &             & $20$     & SP & T/K \\
11.& J$1852-0118$ &           & 0.452  & 3.64  &             &$ 20$     & SP & H\\
12.& J$1903+0925$ &           & 0.357  & 4.10  &             &$ 155$    & SP & M11\\
13.& J$1946+1805$ & B$1944+17$& 0.441  & 3.69  &             &  $7$     & SP & T/K \\\hline
\end{tabular}
\end{minipage}
\end{table*}

\begin{figure}
\begin{center}
\includegraphics{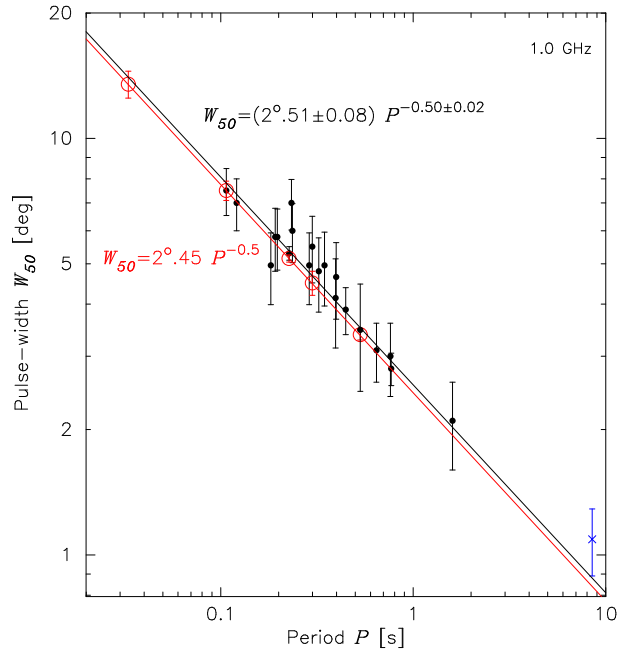}
\caption{Plot of $W_{50}$ vs. $P$ for 21 DP--IP pulsars. Five cases marked by red circles were taken from Table 3 in \citet{r90} and the rest of points represent our new measurements (see Table \ref{tab.1}). The red solid line represents the original fit of \citet{r90}, while the black line represents the fit to all data points. We also added the longest period radio pulsar J$2144-3933$ marked in blue to show that it follows the boundary. However it was not included in the fitting procedure as it did not had the IP emission.  \label{figure.1}}
\end{center}
\end{figure}

Table \ref{tab.1} contains 44 known IP cases divided into double--pole (DP--IP) and single--pole (SP--IP) cases (see \citet{mgr11} for details). This table presents for each case the values of $W_{50}$ predicted by Equation (\ref{eq.1}) with $\alpha =90^{\circ}$ derived by \citet{r90}, as well as those measured by \citet{r90} and measurements made in this paper based on IP studies presented in \citet{mgr11}. Among 31 DP--IP and 13 SP--IP cases 21 and 11 measurements of $W_{50}$, respectively, were possible. We utilized the ATNF\footnote{http://www.atnf.cisiro.au/research/pulsar/psrcat/} data base \citep{manchester05}, where the values of overall half-power widths were tabulated. If needed, the raw data for individual profiles were available for inspection and/or processing\footnote{http://datanet.csiro.au/dap/}. Figure \ref{figure.1} shows the plot $W_{50}$ versus $P$ for DP--IP case only. Red open symbols represent measurements of \citet{r90} presented in her Table 3, while black filled symbols represent our measurements\footnote{We measured or estimated $W_{50}$ of the core components in 21 DP-IP and 11 SP-IP pulsars. In a few cases of DP-IP cases it was possible to obtain the so-called log-polar representation \citep{hankins86} of the profile and measure the core widths quite accurately. However, in most cases the estimate was done by means of the visual inspection of the profile and a reasonable judgement. For example, if the profile was apparently complex but its components could not be resolved we used the following method: we estimated the half-power width of the entire envelope of main pulse or inter-pulse and then we divided the obtained result by an integer following from the complexity of the envelope (mostly 3). In such cases quite conservative errors were given. Our measurements were consistent with those of \citet{r90} in all cases where the comparison was possible. In fitting procedures we used mostly our new measurements except five cases for wich the estimates given by R90 were more accurate than ours.} presented in Table \ref{tab.1} in this paper. We added the case of longest period radio pulsar J$2144-3933$, as its $W_{50}$ value (ATNF database) falls close to the value predicted by \citet{r90} for DP--IP pulsars. Although, the interpulse has not been found in this pulsar, this seems consistent with its very narrow pulse profile (interpulse could be just missed by our line--of--sight). This pulsar was identified as the core profile case \citep{young99} and it had the narrowest pulse profile among all pulsars known at that time.

\begin{figure}
\begin{center}
\includegraphics{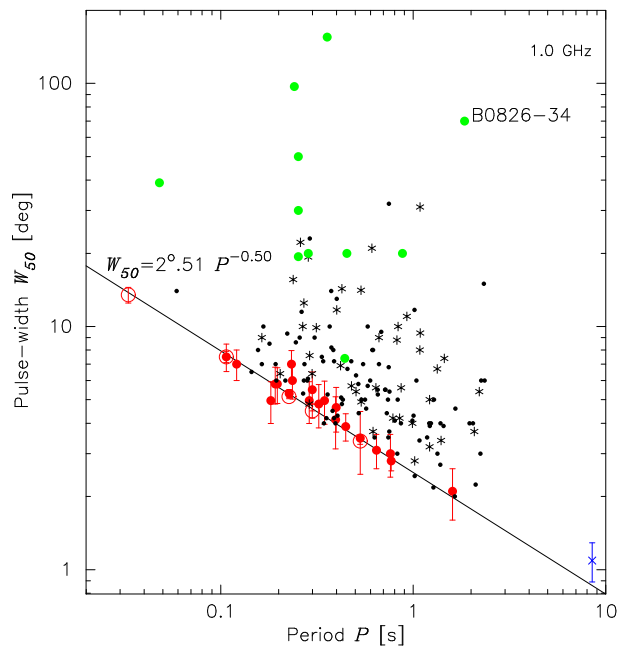}
\caption{Similar as in Figure \ref{figure.1} but with measurements of core widths for a number of other pulsars, including SP--IP cases marked in green. Dots represent the measurements of \citet{r90} while stars represent our measurements. For DP--IP cases marked in red, the open symbols represent measurements of \citet{r90}, while filled symbols represent our measurements. The important case of B$0826-34$ is also marked.  \label{figure.2} }
\end{center}
\end{figure}

The formal fit to all data points (excluding J$2144-3933$) is 
\begin{equation}
      W_{50} = (2^{\circ}.51\pm0^{\circ}.08) \; P^{-0.50\pm0.02},
\label{eq.2}
\end{equation}
which is represented by the black line in Figure \ref{figure.1}.
This is very close to the fit corresponding to the data of R90 (marked in red) $W_{50} = (2^{\circ}.46\pm0^{\circ}.09) \; P^{-0.50\pm0.02}$, represented by red line in Figure \ref{figure.1}. Therefore, the lower bound found by \citet{r90} still appears clearly in our larger data set analysed in this paper. This is illustrated in Figure \ref{figure.2}, which contains $W_{50}$ measurements of DP--IP (red), SP--IP (green) and core components of other pulsars (data of \citet{r90} marked by black dots (Table \ref{tab.c1} in the on-line Appendix \ref{appendix.c}) and our new measurements marked by black stars (Table \ref{tab.c2} in the on-line Appendix \ref{appendix.c}). At least from the first sight the Rankin's Equation (\ref{eq.1}) seems well established. The important case of extremely broad profile pulsar B$0826-34$, with core component width being close to $70^{\circ}$, is marked (see also caption to Figure \ref{figure.5}).

\begin{figure*}
\includegraphics{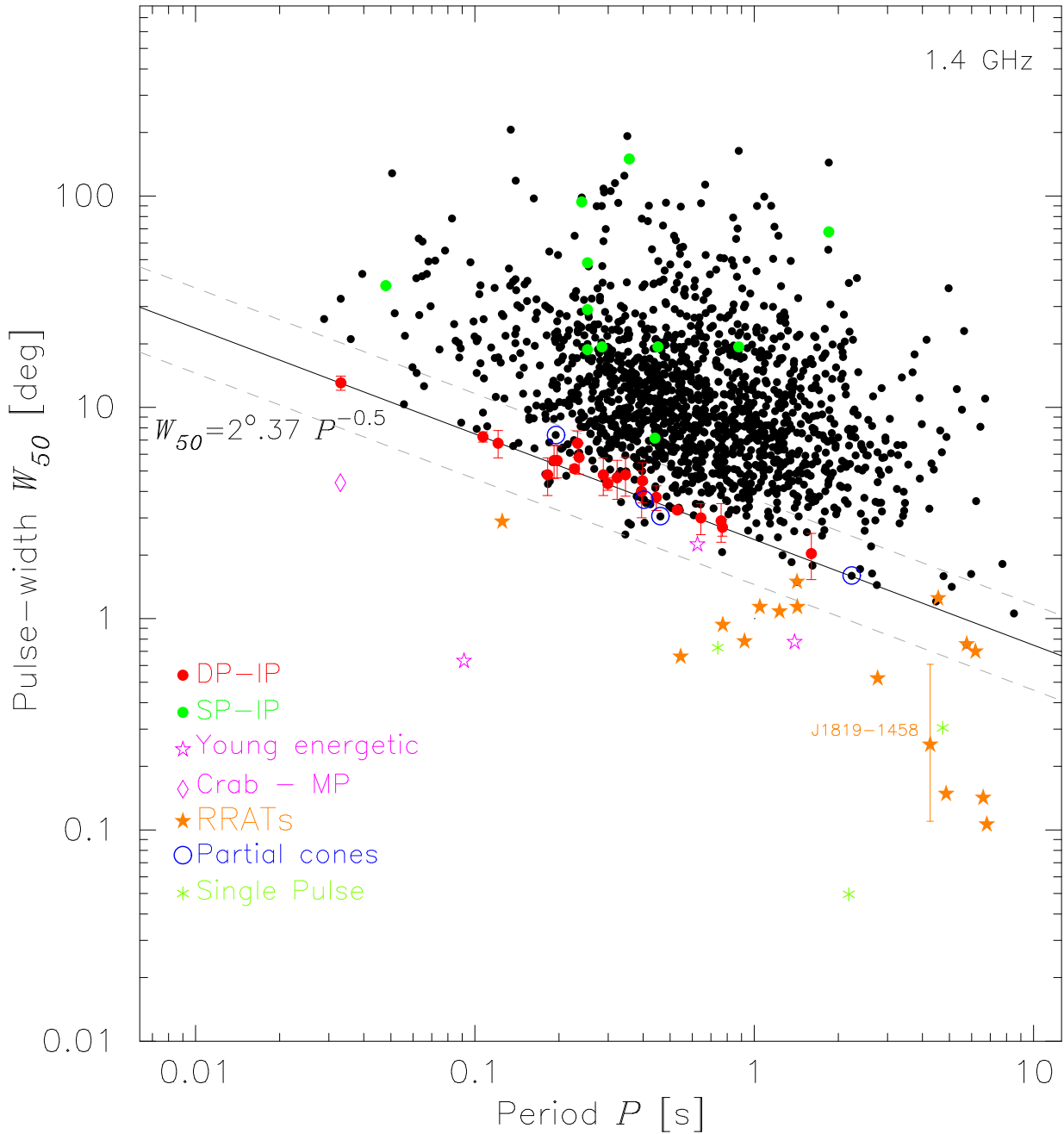}
\vbox to 20mm{\vfil\caption{Plot of pulse--widths $W_{50}$ measured at 1.37 GHz for 1522 normal pulsars (excluding millisecond and other recycled pulsars; see Paper I for details) from ATNF database, with added 21 cases of DP--IP (marked in red) and 11 cases of SP--IP (marked in green) taken from Figure \ref{figure.2} (Table \ref{tab.1}). The black solid line represents the lower limit from Figure \ref{figure.2} rescaled to the frequency 1.37 GHz. The two dashed grey lines are marked to illustrate the scatter of red data points and their errors around the black solid line $W_{50}=2^{\circ}.37\;P^{-0.5}$. The case of J$2144-3933$ is marked by blue cross, while blue circles correspond to pulsars with partial cone profiles. The orange stars represent RRATs, the magenta stars represent young energetic pulsars and green stars represent pulsars discovered as strong single pulses (see Table \ref{tab.2} for the list of special cases).
\label{figure.3}} \vfil}
\end{figure*}

\begin{table}
\begin{center}
  \caption{Special pulsars appearing in Figure \ref{figure.3} near or below the lower boundary line$\dag$.
\label{tab.2}}
  \begin{tabular}{|l|c|c|c|c|}
\hline
\multicolumn{1}{|c|}{Name J} & \multicolumn{1}{|c|}{$P$ [s]} & \multicolumn{1}{|c|}{$W_{50}^{R90}$[deg]}  & \multicolumn{1}{|c|}{$W_{50}$[deg]} &\multicolumn{1}{|c|}{Ref.}\\

\multicolumn{1}{|c|}{}  & \multicolumn{1}{|c|}{}  & \multicolumn{1}{|c|}{1.4GHz}  & \multicolumn{1}{|c|}{1.4GHz} &\multicolumn{1}{|c|}{}\\
\hline
\multicolumn{5}{|c|}{Partial cone}  \\
J$1759-2205$ & 0.461 & 3.61 & 3.05  & MR11 \\
J$1912+2104$ & 2.232 & 1.64 & 1.60  & MR11 \\
J$1917+1353$ & 0.194 & 5.56 & 7.40  & MR11 \\ 
J$1941-2602$ & 0.403 & 3.86 & 3.66  & MR11 \\ \hline
\multicolumn{5}{|c|}{RRATs}            \\        \hline
J$0735-62  $ & 4.865 & 1.11 & 0.15  & KKL11 \\
J$1047-58  $ & 1.231 & 2.21 & 1.08  & KKL11 \\
J$1226-32  $ & 6.193 & 0.98 & 0.70  & KKL11 \\
J$1423-56  $ & 1.427 & 2.05 & 1.14  & KKL11 \\
J$1514-59  $ & 1.046 & 2.40 & 1.14  & KKL11 \\
J$1554-52  $ & 0.125 & 6.93 & 2.88  & KKL11 \\
J$1654-23  $ & 0.545 & 3.32 & 0.66  & KKL11 \\
J$1707-44  $ & 5.764 & 1.02 & 0.76  & KKL11 \\
J$1724-35  $ & 1.422 & 2.05 & 1.49  & KKL11 \\
J$1807-25  $ & 2.764 & 1.47 & 0.52  & KKL11 \\
J$1819-1458$ & 4.263 & 1.19 & 0.25  & HEY11, KHS09 \\
J$1826-1419$ & 0.771 & 2.79 & 0.93  & MLK09 \\
J$1841-14  $ & 6.598 & 0.95 & 0.14  & KKL11 \\
J$1854+03  $ & 4.558 & 1.15 & 1.25  & KKL11 \\
J$1848-12  $ & 6.795 & 0.94 & 0.11  & MLL06 \\
J$1913+1330$ & 0.923 & 2.55 & 0.78  & MLK09 \\  \hline

\multicolumn{5}{|c|}{Young energetic PSRs}\\  \hline
J$0531+2200$ & 0.033 &13.49 & 4.4   & R90  \\
J$1028-5819$ & 0.091 & 8.12 & 0.63  & KJK08  \\
J$1225-6035$ & 0.626 & 3.10 & 2.24  & KBM03  \\
J$1855+0527$ & 1.393 & 2.08 & 0.78  & KEL09  \\   \hline
                                         
\multicolumn{5}{|c|}{Single Pulse } \\      \hline
J$0627+16$   & 2.180 & 1.66 & 0.05  & DCM09  \\
J$1909+06$   & 0.741 & 2.85 & 0.73  & DCM09  \\
J$1946+24$   & 4.729 & 1.13 & 0.31  & DCM09  \\   \hline
\end{tabular}
\end{center}
$\dag$ There were also two recycled pulsars (J$1410-7404$ and J$2129+1210$G) below this line, which we did not show according to strategy adopted in Paper I.
\end{table}

Figure \ref{figure.2} was prepared using $W_{50}$ measurements of the core profile components. However, one can ask the question how would this figure look like if one used all $W_{50}$ measurements available in the ATNF database (without distinctions into core and conal components). Some of the measurements corresponding to profiles dominated by prominent core components would just reproduce the Figure \ref{figure.2}. However, those measurements which correspond to the conal profiles should lie above the Lower Boundary Line $2^{\circ}.5\: P^{-0.5}$ (LBL hereafter), which should still appear as the prominent feature. Figure \ref{figure.3} presents a plot of all 1522 values of $W_{50}$ versus pulsar period, which were available in the ATNF database (excluding millisecond and other recycled pulsars). Apparently, the bulk of the measurements lies above the LBL ($W_{50}=2^{\circ}.37\: P^{-0.5}$ after rescaling to 1.37 GHz). The widths of DP--IP (red symbols with error bars) and SP--IP (green symbols) were added for the reference and comparison with Figure \ref{figure.2}.

It is natural why the absolute majority of points lies above the lower boundary line. Indeed, if these are core components and/or conal components surrounding the core ones, then they must obey the lower boundary limit as discussed in this paper (for geometry of conal components see \citet{r93}, \citet{gks93} and \citet{kramer94}). Therefore, the real question is why there is a small number of points below the LBL (assuming that these measurements are not erroneous)? Taking into account the scatter of red data points and their errors in Figure \ref{figure.3} it is instructive to introduce the Lower Boundary Belt $W_{50}=(2^{\circ}.37^{+1^{\circ}.43}_{-0^{\circ}.87})\: P^{-0.5}$ (LBB hereafter; marked by two grey dashed lines), which should not be mistaken with the formal fit expressed by Equation (\ref{eq.2}). We took every effort to identify points lying below or within the LBB (Table \ref{tab.2}). It is extremely interesting that most of these points were Rotating RAdio Transients (RRATs; \citet{mclaughlin06}) marked by orange stars, or other sporadically emitting pulsars. RRATs are repeating sources with pulsar underlying periodicity, detectable via their extremely strong sporadic single pulses. A good example is the case of J$1819-1458$ marked by the orange error bar in Figure \ref{figure.3}, whose length represents individual measurements of different events \citep{hu11}, with the average value of $W_{50}=0^{\circ}.25$ (Table \ref{tab.2}). Interestingly, the width of the mean profile obtained by folding of large number of individual events of this relatively frequent RRAT is about $10^{\circ}$, well above the LBB. However, the most interesting fact is that the width of the central component of this profile is about $1^{\circ}.4\pm0^{\circ}.4$ (Figure 1. in \citet{karastergiou09}), which fits to the LBL/LBB $W_{50}=2^{\circ}.37\: (4.3)^{-0.5}=1^{\circ}.14$ very well. This strongly suggest that most of data points below the LBB in Figure \ref{figure.3} represent a single events of RRATs activity, whose pulse--widths are much smaller than their underlying pulse windows.

Another small group of three data points below the LBB represents the Single Pulse (SP) detected pulsars (green stars) in the Arecibo PALFA survey \citep{deneva09}. One of them J$0627+16$ has the narrowest ever measured width of $0^{\circ}.05$ (Table \ref{tab.2}), which corresponds to the lowest known duty cycle of about 0.01 percent (compared to the typical value of about 5 percent).

Yet another group of data points below the LBB represents the young energetic PSRs (Table \ref{tab.2}) marked in magenta. It is interesting that the main pulse of the Crab pulsar (marked by diamond) seems to belong to this category as well. It is well known that the precursor of the main pulse in the profile of B$0531+21$ behaves like typical core component, while the main-pulse and the interpulse show rather untypical properties (e.g. \citet{r90}).

We also checked the group of pulsars with partial cone profiles but no single case was found below the LBB in Figure \ref{figure.3}. In the long list of such pulsars recently published by \citet{mitra11} we identified only four cases (Table \ref{tab.2}) lying within the LBB (blue circles). This paradoxical at first sight result means that the opening angles of core and conal beams are about the same.

In summary, the lower boundary with the slope following $P^{-0.5}$ is a prominent feature in the distribution of pulse--width $W_{50}$ versus pulsar period $P$, especially if extended to the concept of narrow lower boundary belt (Figure \ref{figure.3}) with the width determined by scatter and errors of $W_{50}$ measurements of core components in DP--IP pulsars (red data points). There is a small number of data points below the LBB but all of them represent some unusual species, mostly transients and intermittent pulsars, whose sporadic emission is much narrower than their underlying pulse windows. Finally, it is worth noting that the existence of LBL or even LBB is not depending neither on the method$^{3}$ nor on the accuracy of the measurements of core widths in DP--IP pulsars. Indeed, the LBL is still apparent when the red symbols are removed from Figure \ref{figure.3}, as clearly visible in Figure \ref{figure.8}.

\section{Geometry of the core emission}
The period dependent lower boundary in $W_{50}$ pulse--widths revealed  in pulsar data, translates into the period dependent lower limit for the half-power beam widths of the transversal structures in the overall pulsar beam. Geometry of pulsar emission is described by the well known formula
\begin{equation}
W_{50}=4 \; \arcsin\left[\frac{\sin\left[\left(\rho_{50}+\beta\right)/2\right] \;\sin\left[\left(\rho_{50}-\beta\right) /2\right]}{\sin\alpha\ \; sin\left(\alpha+\beta\right)}\right]^{1/2},
\label{eq.3}
\end{equation}
(\citet{gil81,gil84}, Gil, Kijak \& Seiradakis (1993); hereafter GKS93), where $\alpha$ is the inclination angle (of the magnetic axis to the rotation axis), $\beta$ is the impact angle (of the closest approach of the line--of--sight to the magnetic axis) and $\rho_{50}$ is the opening angle of the pulsar beam corresponding to half-power maximum beam width (FWHM). This equation holds generally on the sphere centered on the pulsar and the only assumption about the pulsar beam that has to be satisfied is the phase symmetry about the fiducial plane $\varphi_0$ containing both pulsar axes and the observer's direction (Figure \ref{figure.b1} in the on-line Appendix B). One should realise that the  illustration of circular beam presented in Figure \ref{figure.b1} represents only a cross-section of a certain intensity level of the beam with the pulsar centered celestial hemisphere. It says nothing about the internal structure of the beam, neither the angular nor the radial intensity distribution. The core pulsar beam can be modelled by a gaussian distribution of the mean intensity $I$, symmetric about the magnetic axis (beam axis). Thus, it can be described by the function
\begin{equation}
      I=I_0 \; \textnormal{exp}[-k(\rho/\rho_0)^2],
\label{eq.4}
\end{equation}
where $I_0=I(\beta=0)=I_{max}$, $I(\rho_0)=I_0 \; \textnormal{exp}(-k)=I_{min}$ and $k$ is an arbitrary steepness parameter. For a certain impact angle $0 << \beta < \rho$ the observer will detect a local intensity maximum
\[
      I_m(\beta) = I_0 \; \textnormal{exp}[-k(\beta/\rho_0)^2]
\]
and thus we can write Equation (\ref{eq.4}) in the form 
\begin{equation}
      I(\rho,\beta)=I_m(\beta) \; \textnormal{exp} \left[-\frac{k}{\rho_0^2}(\rho^2-\beta^2)\right],
\label{eq.5}
\end{equation}
where $I_m(\beta)$ is the local peak intensity which can be determined from observations. R90 used slightly different approach and assumed that the angular distribution of the core pulsar beam is well approximated by a bivariate gaussian function
\[
      I(\beta,\varphi) \propto \; \textnormal{exp} -[(\beta-\beta_0)^2/2\rho_{\beta}^2 + (\varphi - \varphi_0)^2/2\rho_{\varphi}^2],
\]
whose breadth (FWHM) is invariant along any set of parallel paths. Thus, she concludes, ''the angular width of core beams can only be weakly dependent on the centrality of traverse and the effect of the impact angle $\beta$ on the observed width of the core components may be quite weak''. However, one should realise that approximation described by the above equation is a true representation of gaussian intensity distribution of the pulsar beam only in a special case when $\varphi\sim\rho$, that is $\beta\approx0^{\circ}$ and $\alpha\sim90^{\circ}$ (Equation \ref{eq.3}). Only then the above function $I(\beta,\varphi)$ is equivalent to $I(\rho,\beta)$ expressed by our exact Equation (\ref{eq.5}). In general, although the projection of the line--of--sight cut through the gaussian beam is indeed independent of the impact angle $\beta$, the observed pulse--width (not a beam width) $W=2\varphi$ may depend on both $\alpha$ and $\beta$. To examine this problem exactly we will use Equation (\ref{eq.3}) equipped with the relationship
\begin{equation}
      \rho_{50}(\beta) = \left(\rho_{50}^2 + \beta^2\right)^{1/2}
\label{eq.6}
\end{equation}
valid only for a gaussian beam shape (see Figure \ref{figure.b2} for illustration). Here $\rho_{50} = \rho_{50}(\beta=0)$ is the half-power beam width, while $\rho_{50}(\beta)$ is the opening angle of pulsar beam measured at FWHM level by the observer with the impact angle $\beta$. Following Figures \ref{figure.1} and \ref{figure.2} it is quite reasonable to adopt for the half-power width of the core beam
\begin{equation}
    \rho_{50}=1^{\circ}.2\: P^{-1/2},
\label{eq.7}
\end{equation}
which seems quite well established observationally\footnote{The exact value that should appear in Equation (\ref{eq.7}) is just a half of the 2.45 factor from Equation (\ref{eq.1}). Of course such a four significant figures accuracy is not important, we nevertheless used it in our numerical calculations.}. The natural $P^{-1/2}$ factor undoubtedly means that the radiation associated with $W_{50}$ of core components is emitted tangentially to dipolar magnetic field lines, but the value of the numerical factor close to $1^{\circ}.2$ is not yet clear (see Conclusion section).

\begin{figure}
\begin{center}
\includegraphics{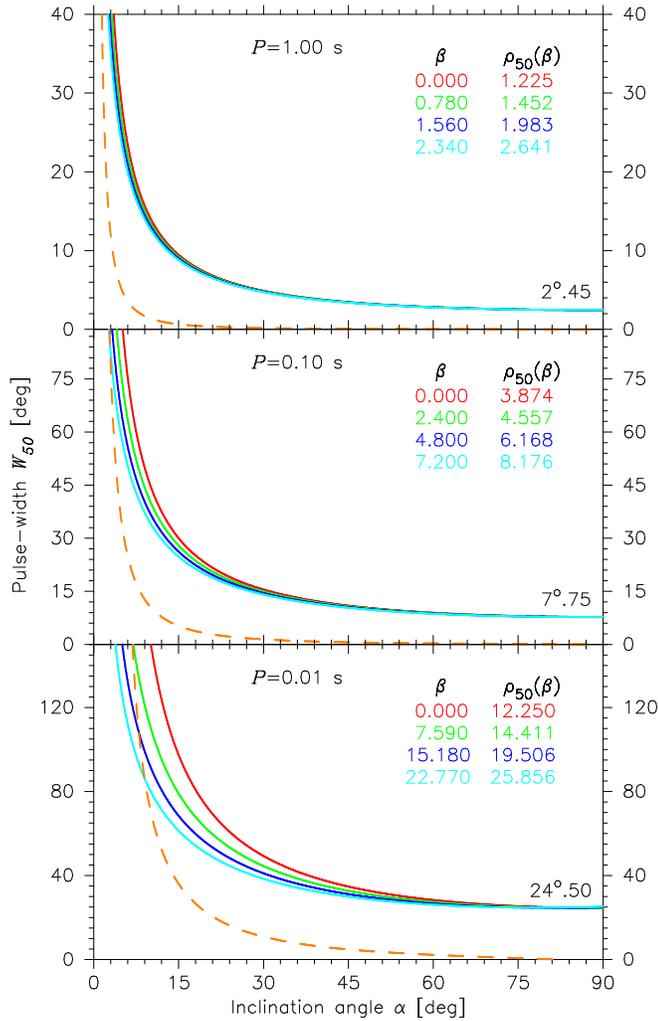}
\caption{Calculations of $W_{50}$ vs. $\alpha$ using Equations (\ref{eq.3}), (\ref{eq.6}) and (\ref{eq.7}). Different panels correspond to different periods $P$, while four different colours of solid lines correspond to different values of $\beta$ and $\rho_{50}(\beta)$. The asymptotic values $2\rho_{50}(0)$ are marked in the lower right corner of each panel. The dashed orange lines describe the maximum difference between the solid colour lines. \label{figure.4}}
\end{center}
\end{figure}

\begin{figure}
\begin{center}
\includegraphics{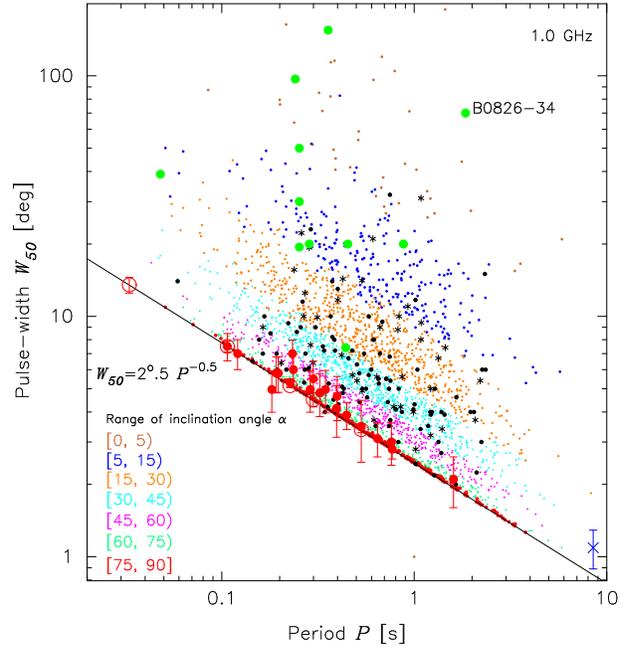}
\caption{The results of the Monte Carlo simulations of pulse--widths superposed on the Figure \ref{figure.2}. Six different ranges of the inclination angle $\alpha$ are marked by different colours. The DP--IP and SP--IP cases are marked by red and green symbols, respectively. The opened red symbols represent measurements of R90, while filled red symbols represent our new measurements (see Table \ref{tab.1}). The solid line represents the fit to red marked data points expressed by Equation (\ref{eq.2}). For a very wide profile pulsar B$0826-34$ with SP--IP the simulations indicate the value of the inclination angle $\alpha<1^{\circ}$, which is consistent with an independent estimate of \citet{esamdin05} $\alpha=0^{\circ}.5$ for this almost aligned rotator. \label{figure.5}} 
\end{center}
\end{figure}

The results of calculations of $W_{50}(\alpha,\beta,\rho(\beta))$ for different values of the pulsar period $P$ are presented in Figure \ref{figure.4}. Three different panels correspond to different periods $P$ equal to 1, 0.1 and 0.01 seconds, respectively. Four solid lines in each panel correspond to different values of the impact $\beta$ and opening $\rho_{50}(\beta)$ angles marked by red, green, dark blue and light blue colours (the latter corresponds to the value of about $0.9\: \rho_{50}(\beta)$). The dashed orange line represents the maximum difference between the solid lines (red and light blue). As one can see for $P=1$s a significant residual occur only for very small inclination angles $\alpha<5^{\circ}$ (almost aligned rotators). For shorter periods the residuals are larger. However, even for 10 millisecond periods the errors made due to neglecting $\beta$ are negligible for inclinations angles $\alpha > 30^{\circ}$. Thus, one can generally state that neglecting $\beta$ in estimating the core width is quite well justified by results of strict calculations, especially for larger inclinations angles. For orthogonal rotators (DP--IP) it can be treated as an exact result. Thus, Equation (\ref{eq.1}) proposed by \citet{r90} should be regarded as a very good tool for estimating the inclination angles using the core pulse--widths. In Figure \ref{figure.5} we superposed on the data from Figure \ref{figure.2} the results of simulations of $W_{50}$ performed using a technique developed in Paper I applied to core emission (see on-line Appendix \ref{appendix.a}). Although these were strict calculations, including simulations of both inclination angle $\alpha$ (different ranges of this angle are marked with different colours) and impact angle $\beta$, their results are fully consistent with Equation (\ref{eq.1}). This equation was derived  by \citet{r90} assuming $\beta=0$, based on the claim that the core width is almost completely insensitive of the value of the impact angle $\beta$. Pulsars with interpulses corresponding to DP--IP and SP--IP models are marked by red and green symbols, respectively. All DP--IP cases lie on the boundary line (to within error bars), as expected. A very interesting case of almost aligned rotator PSR B$0826-34$ is marked in Figure \ref{figure.5}. The inclination angle of this extremely broad profile pulsar was estimated as having very small value $\alpha < 5^{\circ}$ \citep{gupta04} or even $\alpha<1^{\circ}$ \citep{esamdin05}, which is consistent with results of our simulations.

\section{Inclination angles and pulse--widths}
\begin{figure}
\begin{center}
\includegraphics{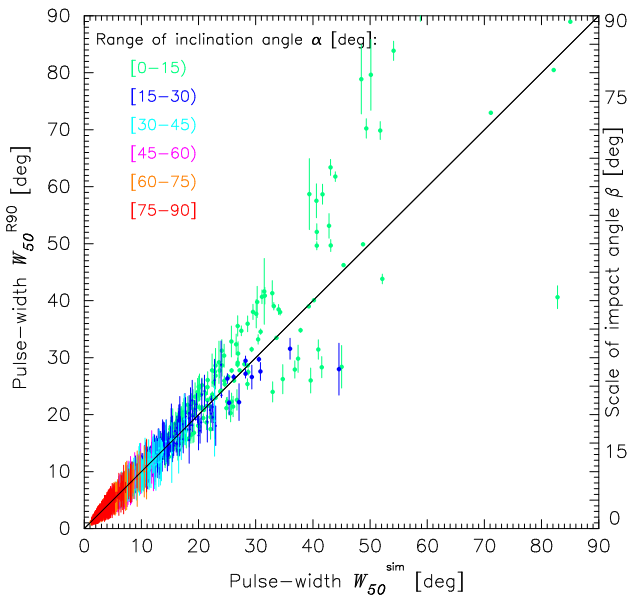}
\caption{Comparison of $W_{50}$ values obtained by means of Equation (\ref{eq.1}) and simulations based on Equations (\ref{eq.3}), (\ref{eq.6}) and (\ref{eq.7}). Six different ranges of $\alpha$ values are marked by different colours, while values of $\beta$ are represented by the length of vertical bars with the same scale as that of the right-hand side ordinate axis. \label{figure.6}}
\end{center}
\end{figure}

As demonstrated in Figure 4 using Monte Carlo simulations we can calculate the pulse--widths $W_{50}$ of the core emission for large population of pulsars. These calculations are based on Equations (\ref{eq.1}) and (\ref{eq.3}), with $\alpha$ and $\beta$ values drawn from suitable distributions described in Paper I (see on-line Appendix A). On the other hand we can calculate values of $W_{50}$ using Rankin's Equation (\ref{eq.1}), with $\alpha$ and $P$ values generated by Monte Carlo simulations. Figure \ref{figure.6} presents comparison of $W_{50}$ values obtained by these two different methods. Six different ranges of $\alpha$ values are marked by different colours, while values of $\beta$ are represented by the length of vertical bars with the same scale as that of the ordinate axis. Of course, the correlation is perfect for very small $\beta\sim0^{\circ}$. As one can see the correlation is also very good for larger $\beta$ (corresponding to longer vertical bars), provided that inclination angles are not too small ($\alpha > 20^{\circ}$). The points above and below the perfect correlation line correspond to positive and negative values of the impact angle $\beta$, respectively. Generally one can say, that except of cases of almost aligned rotators both methods give very similar results. This further confirms usefulness of Rankin's method expressed by her simple Equation (\ref{eq.1}).

\begin{figure}
\begin{center}
\includegraphics{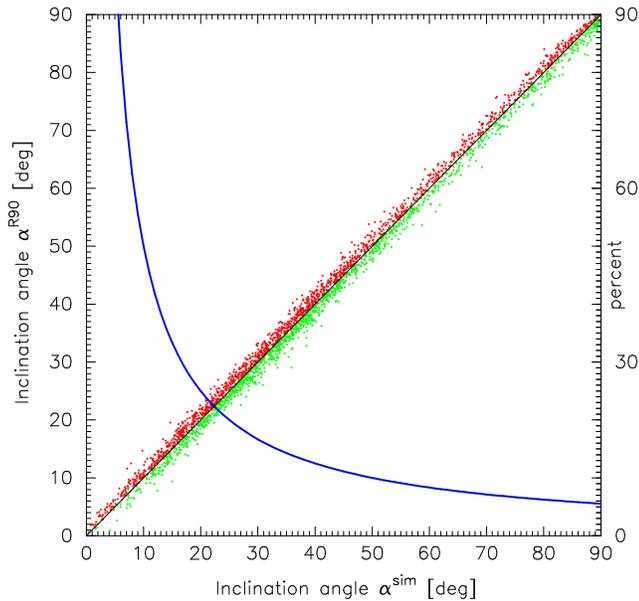}
\caption{Scatter plot of inclination angles $\alpha$ obtained using Rankin's method (Equation (\ref{eq.1})) and Monte Carlo simulated values. Red and green colours on the scatter plot correspond to positive and negative values of the impact angle $\beta$, respectively. The blue solid line represents the maximum relative difference between two methods (in percents according to the right--hand side ordinate scale). \label{figure.7}}
\end{center}
\end{figure}

Finally, we can make a direct comparison of Rankin's method of inclination angle estimations represented by Equation (\ref{eq.1}) with strict calculations based on Equations (\ref{eq.3}), (\ref{eq.6}) and (\ref{eq.7}). This comparison is presented in a form of scatter plot in Figure \ref{figure.7}. The ordinate (vertical) axis represents $\alpha$ calculated from Equation (\ref{eq.1}) with $W_{50}$ calculated from Equations (\ref{eq.3}), (\ref{eq.6}) and (\ref{eq.7}), where $\alpha$, $\beta$ and $P$ were taken from Monte Carlo simulations, while the abscissa (horizontal) axis represents directly simulated values of $\alpha$. Red and green colours correspond to positive and negative values of $\beta$ (for details of simulations and detection conditions see Appendix \ref{appendix.a}). The blue line represents the maximum relative difference between scattered points (in the same scale as for $\alpha$ axes but expressed in percents), thus it describes an accuracy of determination of $\alpha$. As one can see, for almost orthogonal rotators the difference between the two methods is about 10 percent. However, for almost aligned rotators the error can reach or even exceed 100 percent. For intermediate values of $\alpha$ the accuracy is in the range of 20 -- 30 percent. This means that Rankin's method of estimating the inclination angle from the width of the core component is quite good an approximation, provided that these values are not too small. In particular, this method should not be applied to extremely broad profile pulsars for which one can expect very small inclination angles $\alpha\le5^{\circ}$.

\section{Conclusions}

\begin{figure}
\begin{center}
\includegraphics{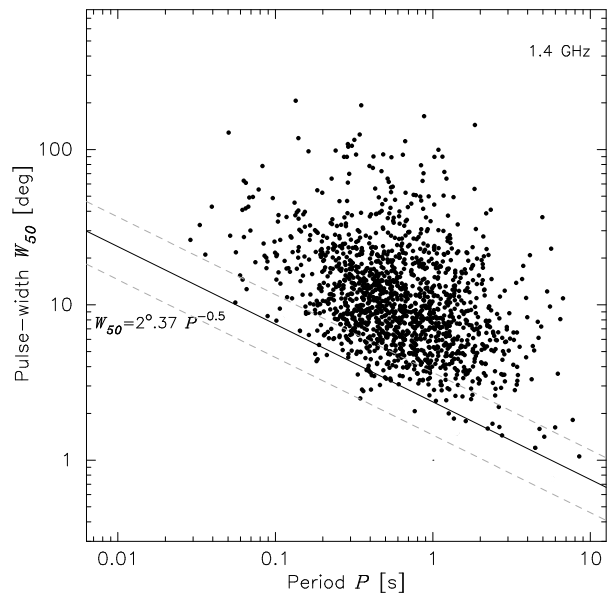}
\caption{Plot of pulse--width $W_{50}$ versus period $P$ for 1495 normal pulsars from the ATNF database. \label{figure.8}}
\end{center}
\end{figure}
The most important result of our paper is a clear confirmation of the lower limit in the distribution of half-power pulse--width $W_{50}$ in core components of pulsar profiles when presented versus the pulsar period $P$ using much bigger database than that used originally by \citet{r90}. As illustrated in Figures \ref{figure.1}, \ref{figure.2}, \ref{figure.3} and \ref{figure.8} this limit is represented by the line $W_{50}=2^{\circ}.5\: P^{-0.5}$ at 1 GHz and $W_{50}=2^{\circ}.37\: P^{-0.5}$ at 1.4 GHz. It seems that more appropriate description of the lower boundary limit is the lower boundary belt (LBB) $W_{50}=(2^{\circ}.37^{+1^{\circ}.43}_{-0^{\circ}.87})\: P^{-0.5}$, which accounts for scatter and  errors in $W_{50}$ measurements. Strictly speaking, the widths of the LBB was determined by the scatter and errors of DP-IP cases marked as red symbols in Figure 3. However, the normal data points seem to populate the belt marked in Figure 8 in similar manner as the DP-IP cases do in Figure 3. The concept of LBB is based not only on core widths of DP--IP pulsars (like in \citet{r90}) but here it is based on all $W_{50}$ measurements of normal pulsars available in the ATNF database, without distinctions into core and conal components, of course. This is dramatically illustrated in Figure \ref{figure.8} containing 1495 data points. This figure is a reproduction of Figure \ref{figure.3} with all special points described in the legend omitted. Apparently, the existence of the lower boundary limit in the pulse--width measurements discovered by Rankin (1990) does not depend neither on the method nor on the errors of measurements of the core widths in DP-IP pulsars. The unweighted formal fit to the points within LBB (between dashed lines) is $W_{50}=(2^{\circ}.86\pm0^{\circ}.07)\: P^{-0.51\pm0.02}$.

The existence of the lower bound in the distribution of $W_{50}$ versus pulsar period $P$ is very intriguing and requires solid and  convincing explanation. The exponent $-0.5$ is evidently related to the geometry of dipolar field lines in the radio emission region, while the value of the numerical factor about $2^{\circ}.5$ is a little more mysterious. \citet{r90} argued that the core component widths are intimately related to the polar cap geometry at the stellar surface, whose bundle of the last open dipolar field lines have an angular extent (opening angle $2\rho$) very close to $2^{\circ}.45 \: P^{-1/2}$. Although this natural and very appealing explanation is commonly accepted, it is not free from theoretical and interpretational problems. First of all, the coherent radio emission cannot originate at or very near the polar cap, because there are no plasma instabilities available in this region. Recently, a modern version of this idea is discussed, with arguments that the core emission originates below the conal one but the former is emitted well above the polar cap surface. For example, \citet{mitra07} argued that in PSR B$0329+54$ the emission of its core component originates over a range of altitudes (up to about 100 km) above the star surface (see also \citet{gil91}). This broadens the required opening angle by a factor of about square root of the normalised emission altitude, that is about 3. If this is a case then one has to conclude that the core radio emission is related to an innermost subset of the open dipolar magnetic field lines. The problem therefore is to identify structures within the overall pulsar beam with an angular extent corresponding to the observed lower boundary limit. We plan to devote to this problem the forthcoming Paper III of this series.

The interpretational problems mentioned above do not affect the approximate method of estimating the inclination angle $\alpha$ in any pulsar with a core component proposed by \citet{r90}. We have shown by means of Monte Carlo simulations based on the strict geometrical calculations that this method offers quite good an approximation with the accuracy of 5, 10, 20 and 50 percent for $\alpha\sim90^{\circ}$, $\alpha\sim50^{\circ}$, $\alpha\sim25^{\circ}$ and $\alpha\sim10^{\circ}$, respectively (Figure \ref{figure.7}). This method cannot be used in almost aligned cases with very small values of $\alpha$.

\section*{Acknowledgments}
We thank Prof. R.N. Manchester for invitation of K.M. to Australia, support and access to the ATNF pulsar data. This paper includes archived data obtained through the Australia Telescope Online Archive and the CSIRO Data Access Portal (http://datanet.csiro.au/dap/). We are grateful to Dr. D. Mitra for critical reading of the manuscript and insightful comments.

\bsp

\newpage
\begin{center}
    \LARGE{ON--LINE MATERIALS}  \label{sec.appendix.online}
\end{center}
\appendix
\section{Monte Carlo simulations of the core emission} \label{appendix.a}
To simulate the core pulsar emission we use the technique described in our Paper I and adapt it to the specifics of coral beams. We will simulate only normal pulsars, that is excluding all millisecond, binary and other recycled pulsars. In addition to Equation (\ref{eq.3}) we will use Equations (\ref{eq.6}) and (\ref{eq.7}) valid exclusively for gaussian core beam. Following a technique used in Paper I we will usually make a number of subsequent steps. 
\begin{enumerate}
\item[1.] Generate the pulsar period $P$ as a random number with the parent probability density function $f(P)$ corresponding to log--normal distribution function (Equation (8) in Paper I, which seem to best suited statistically for representing the true parent distribution of pulsar periods).
\item[2.] Generate the inclination angle $\alpha$ as a random number with the parent probability density function $f(\alpha)$ corresponding to Equation (7) in Paper I, which seems to be best suited for reproducing the right ratios of interpulse cases in pulsar population.
\item[3.] Generate the observer angle $\xi=\alpha+\beta$ as a random number with the parent probability density function $f(\xi) = sin \xi$.
\item[4.] Calculate the impact angle $\beta = \xi - \alpha$.
\item[5.] Calculate the opening angle $\rho_{50}$ for the 50 percent of the maximum intensity of the core component according to Equation (\ref{eq.6}).
\item[6.] Check the detection condition: $\rho_{50} > 3^{\circ} P^{-1/2}$. To the next steps only detected pulsars are taken. The numerical factor $3^{\circ}$ for low intensity edge of gaussian core beams was adopted arbitrarily, but its value influences only the number of detected pulsars and has no important meaning for conclusions of this paper.
\item[7.] Calculate the pulse--width of the core component at the 50 percent of the maximum intensity level:
\item[a)]$W_{50}^{R90}$ according to the Equation (\ref{eq.1}) using simulated $\alpha$ and $P$ values.
\item[b)]$W_{50}^{sim}$ according to the Equation (\ref{eq.3}), where $\rho_{50}$ is calculated from Equations (\ref{eq.6}) and (\ref{eq.7}).
\item[8.] Calculate $\alpha^{R90}$ angle according to Equation (\ref{eq.1}), where $W_{50}=W_{50}^{sim}$.
\end{enumerate}

For more technical details on Monte Carlo simulations of pulsar radio emission see Paper I.

\section{Geometry of pulsar radiation} \label{appendix.b}
\begin{figure}
\begin{center}
\includegraphics[width=8cm,height=6cm,angle=0]{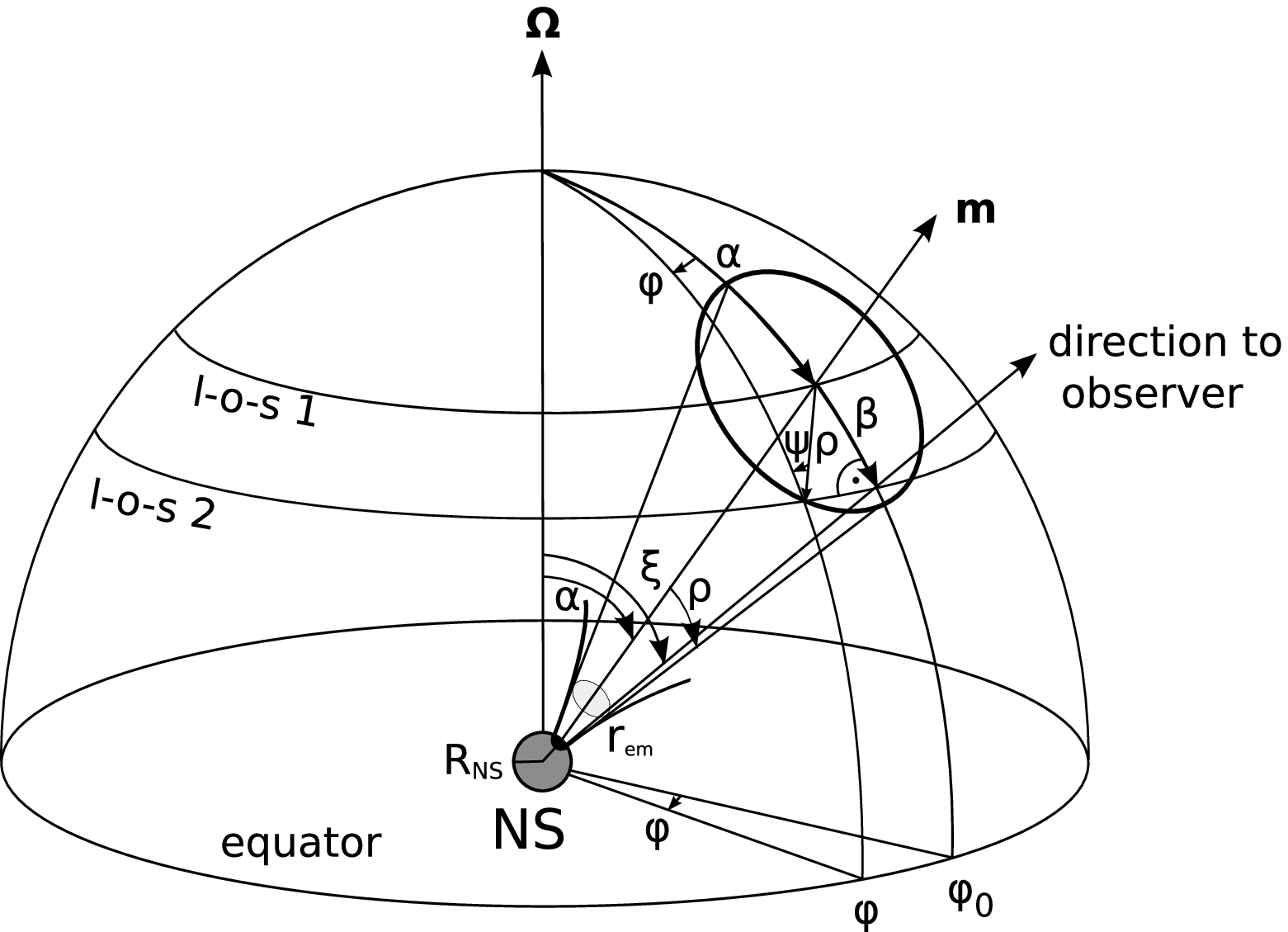}
\caption{Geometry of the pulsar radiation. Schematic picture of the celestial hemisphere centered on the pulsar. The fiducial plane $\phi=0$ is set by the rotation $\mathbf{\Omega}$ and the magnetic $\mathbf{m}$ axes. The following angles are marked: longitudinal phase angle $\phi$ measured from the fiducial plane $\phi=0$, the inclination angle $\alpha$, the impact angle $\beta$, the observer angle $\xi=\alpha+\beta$, the opening angle of the beam of radiation $\rho$ and the polarisation angle $\psi$.
\label{figure.b1}}
\end{center}
\end{figure}

Figure \ref{figure.b1} represents a general view of the geometry of pulsar radiation emitted tangentially to dipolar magnetic field lines and intersected with the celestial sphere centred on the neutron star. 

\begin{figure}
\begin{center}
\includegraphics{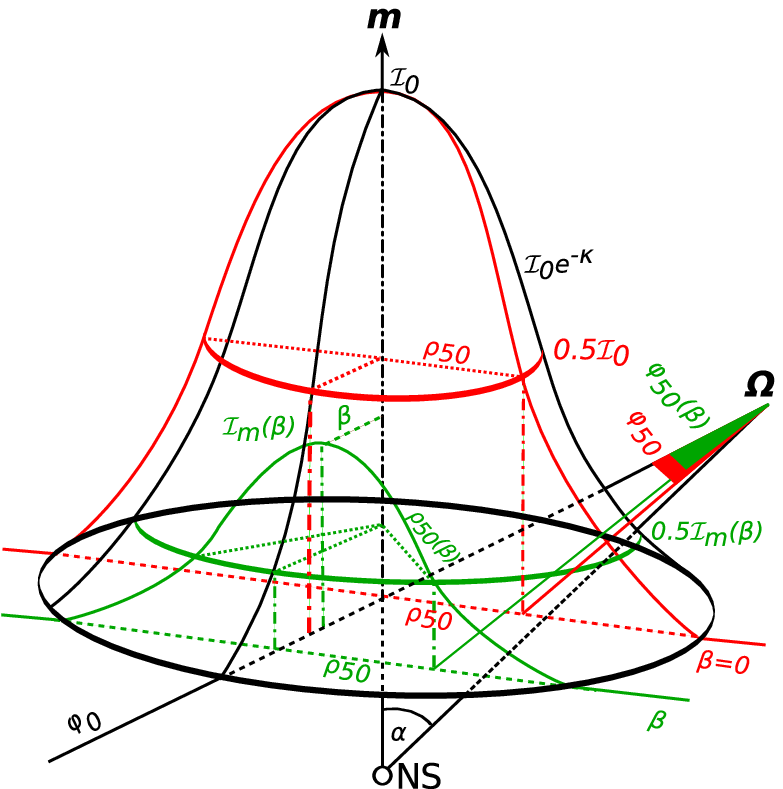}
\caption{The scheme of the gaussian-shaped distribution of the radiation intensity on the pulsar beam centred on the magnetic axis $\mathbf{m}$ inclined to the rotation axis $\mathbf{\Omega}$ at $\alpha$ angle. Two different trajectories are marked: central (red) and side (green), corresponding to the impact angle $\beta=0$ and arbitrary $\beta$, respectively. In both cases the 50 percent level of the local maximum is showed. It is easy to see that in both cases the half-width of the appropriate 2-D Gauss function is the same ($\rho_{50}$). However, the corresponding widths of the observed radiation are different $\phi_{50}(\beta)<\phi_{50}(0)$. Note that $\rho_{50}^2(\beta)=\rho_{50}^2+\beta^2$ for gaussian beam.
\label{figure.b2}}
\end{center}
\end{figure}

\begin{figure}
\begin{center}
\includegraphics{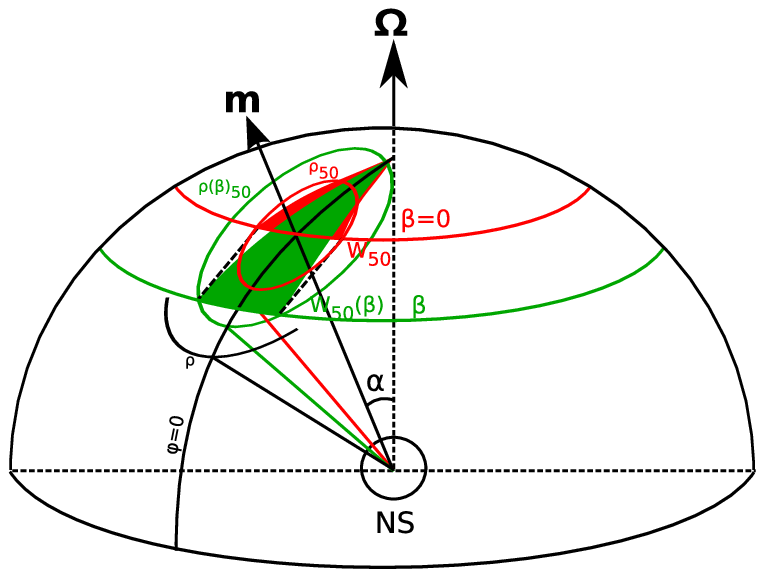}
\caption{The gaussian--shaped pulsar beam presented in Figure \ref{figure.b2} is projected on the spherical surface centered on the NS. Note that $W_{50}(\beta)<W_{50}$, where $W_{50}(\beta)=2\phi_{50}(\beta)$ and $W_{50}=2\phi_{50}$.
\label{figure.b3}}
\end{center}
\end{figure}

Figure \ref{figure.b2} presents 3-D geometry of the gaussian core pulsar beam, while Figure \ref{figure.b3} presents intersection of this gaussian beam with the pulsar centered celestial sphere.

\section{New database of pulse--width $W_{50}$ of core components} \label{appendix.c}
In this section we present a number of tables containing measurements of FWHM pulse--widths of core components that were possible to make judging from published profiles. The references to appropriate papers are given.

\onecolumn

\begin{longtable}{|l|l|l|l||l|l|l|l|}
\caption{Pulse--widths of the core components at the 50\% maximum intensity level taken from Rankin (1990).
\label{tab.c1} }\\

\hline 

\multicolumn {8}{|c|}{\textbf{Pulsars with interpulse}}\\ \hline
\multicolumn{1}{|l|}{No.}  &  
\multicolumn{1}{c|}{Name B}  &  
\multicolumn{1}{c|}{$W_{50}$ [$^\circ$]}    &    \multicolumn{1}{c||}{$P$ [s]}  & 

\multicolumn{1}{c|}{No.}  &  
\multicolumn{1}{c|}{Name B}  &  
\multicolumn{1}{c|}{$W_{50}$ [$^\circ$]}    &    \multicolumn{1}{c|}{$P$ [s]}   \\ \hline\hline

1 & 0531+21p & 13.5 & 0.0331 &  6 & 1055-52  & 5.3  & 0.197  \\
2 & 0531+21  & 4.4  & 0.0331 &  7 & 1702-19  & 4.5  & 0.2989 \\
3 & 0823+26  & 3.38 & 0.5306 &  8 & 1822-09  & 2.8  & 0.7689 \\
4 & 0826-34  & 40   & 1.8489 &  9 & 1929+10  & 5.15 & 0.2265 \\
5 & 0906-49  & 7.5  & 0.1067 &  10 & 1736-29 & --   & 0.3228 \\\hline\hline

\multicolumn {8}{|c|}{\textbf{Pulsars without interpulse}}\\ \hline \multicolumn {8}{|c|}{Pulsars S\textsubscript{t} type (Single)}   \\ \hline
1 & 0105+65 & 5.  &  1.2836 & 24 &  1844-04 & 6.  &  0.5977 \\
2 & 0136+57 & 7.  &  0.2724 & 25 &  1859+03 & 5.25&  0.6554 \\
3 & 0154+61 & 6.  &  2.3517 & 26 &  1900+05 & 6.8 &  0.7465 \\
4 & 0355+54 & 8.  &  0.1563 & 27 &  1900+01 & 4.4 &  0.7293 \\
5 & 0540+23 & 8.6 &  0.2459 & 28 &  1907+02 & 4.1 &  0.9898 \\
6 & 0611+22 & 7.5 &  0.3349 & 29 &  1907+10 & 6.1 &  0.2836 \\
7 & 0626+24 & 7.2 &  0.4766 & 30 &  1907-03 & 6.3 &  0.5046 \\
8 & 0740-28 & 10. &  0.1667 & 31 &  1911-04 & 3.  &  0.8259 \\
9 & 0835-41 & 3.7 &  0.7516 & 32 &  1913+10 & 4.3 &  0.4045 \\
10 & 0940-55 & 7.  &  0.6643 & 33&  1914+13 & 6.  &  0.2818 \\
11 & 0942-13 & 4.6 &  0.5702 & 34&  1915+13 & 6.  &  0.1946 \\
12 & 0959-54 & 3.9 &  1.4365 & 35&  1924+16 & 5.7 &  0.5798 \\
13 & 1154-62 & 13. &  0.4005 & 36&  1929+20 & 5.3 &  0.2682 \\
14 & 1240-64 & 7.2 &  0.3884 & 37&  1930+22 & 6.5 &  0.1444 \\
15 & 1449-64 & 8.5 &  0.1794 & 38&  1933+16 & 5.25&  0.3587 \\
16 & 1556-44 & 9.0 &  0.2570 & 39&  1946+35 & 5.5 &  0.7173 \\
17 & 1557-50 & 7.0 &  0.1926 & 40&  1953+29 & 40  &  0.0061 \\
18 & 1641-45 & 6.7 &  0.4550 & 41&  1953+50 & 5.8 &  0.5189 \\
19 & 1642-03 & 4.2 &  0.3876 & 42&  2002+31 & 2.24&  2.1112 \\
20 & 1706-16 & 5.5 &  0.6530 & 43&  2053+36 & 9.35&  0.2215 \\
21 & 1749-28 & 5.  &  0.5625 & 44&  2113+14 & 7.4 &  0.4401 \\
22 & 1839+09 & 4.5 &  0.3813 & 45&  2217+47 & 5.  &  0.5384 \\
23 & 1842+14 & 8.2 &  0.3754 & 46&  2255+58 & 10. &  0.3682 \\
\hline\hline
 \multicolumn {8}{|c|}{Pulsars T type (Triple)}   \\ \hline
1&  0329+54 & 5.5 &  0.7145 &  20 &    1907+00 & 2.43&  1.0169 \\
2&  0450+55 & 7.2 &  0.3407 &  21 &    1907+03 & 15. &  2.3302 \\
3&  0450-18 & 7.  &  0.5489 &  22 &    1911+13 & 4.4 &  0.5214 \\
4&  0656+14 & 8.  &  0.3848 &  23 &    1913+16 & 14. &  0.0590 \\
5&  0736-40 & 14. &  0.3749 &  24 &    1914+09 & 6.  &  0.2702 \\
6&  0919+06 & 5.  &  0.4306 &  25 &    1916+14 & 3.5 &  1.1810 \\
7&  1112+50 & 4.6 &  1.6564 &  26 &    1917+00 & 2.18&  1.2722 \\
8&  1221-63 & 6.  &  0.2164 &  27 &    1919+14 & 8.  &  0.6181 \\
9&  1451-68 & 11.5&  0.2633 &  28 &    1920+21 & 3.4 &  1.0779 \\
10&  1508+55 & 5.  &  0.7396 & 29 &    1926+18 & 3.9 &  1.2204 \\
11&  1541+09 & 32. &  0.7484 & 30 &    1952+29 & 4.8 &  0.4266 \\
12&  1558-50 & 4.  &  0.8642 & 31 &    2003-08 & 10. &  0.5808 \\
13&  1604-00 & 5.1 &  0.4218 & 32 &    2020+28 & 4.  &  0.3434 \\
14&  1700-32 & 4.  &  1.2117 & 33 &    2045-16 & 4.0 &  1.9615 \\
15&  1727-47 & 5.  &  0.8298 & 34 &    2111+46 & 11.7&  1.0146 \\
16&  1747-46 & 3.5 &  0.7423 & 35 &    2224+65 & 9.5 &  0.6825 \\
17&  1804-08 & 7.  &  0.1637 & 36 &    2319+60 & 6.  &  2.2564 \\
18&  1821+05 & 5.4 &  0.7529 & 37 &    2327-20 & 2.  &  1.6436 \\
19&  1826-17 & 9.  &  0.3071 &  &            &     &         \\
\hline\newpage

\multicolumn {8}{l}{Tab. \ref{tab.c1} cont.}\\\hline
\multicolumn{1}{|l|}{No.}  &  
\multicolumn{1}{c|}{Name B}  &  
\multicolumn{1}{c|}{$W_{50}$ [$^\circ$]}    &    \multicolumn{1}{c||}{$P$ [s]}  & 

\multicolumn{1}{c|}{No.}  &  
\multicolumn{1}{c|}{Name B}  &  
\multicolumn{1}{c|}{$W_{50}$ [$^\circ$]}    &    \multicolumn{1}{c|}{$P$ [s]}   \\ \hline\hline

\multicolumn {8}{|c|}{Pulsars M type (Multiple)}   \\ \hline
1 & 0523+11 & 4.2 &  0.3544 & 8  &  1831-04 & 23. &  0.2901 \\
2 & 0621-04 & 6.  &  1.0390 & 9  &  1857-26 & 8.  &  0.6122 \\
3 & 1039-19 & 4.  &  1.3863 & 10  &  1905+39 & 4.  &  1.2357 \\
4 & 1237+25 & 2.7 &  1.3824 & 11  &  1910+20 & 3.  &  2.2329 \\
5 & 1737+13 & 4.1 &  0.8030 & 12  &  1919+21 & 3.  &  1.3373 \\
6 & 1738-08 & 4.  &  2.0430 & 13  &  2028+22 & 3.5 &  0.6305 \\
7 & 1745-12 & 4.  &  0.3941 & 14  &  2210+29 & 4.3 &  1.0045 \\ \hline
\end{longtable}

\begin{longtable}{|c|c|c|c|}
\caption{New pulse--widths of the core components $W_{50}$ obtained in this paper. The pulse--widths $W$ at different frequencies $f$ were corrected to about 1 GHz according to $W\propto\rho\propto r^{1/2}(f)\propto f^{-0.1}$ (for details see footnote 4 in \citet{mgr11}).
\label{tab.c2}}\\\hline
\multicolumn{4}{|c|}{WEISBERG 1999 -- 1.418 GHz}\\ \hline
Name B & $P$ & $W_{50}$ (1.418 GHz)& $W_{50}^{core}$ (1 GHz)\\ 
& [s] & [$^{\circ}$] & [$^{\circ}$]  \\ \hline
0609+37 & 0.298 & 6.2 & 6.4 \\
0940+16 & 1.087 & 9.1 & 9.4\\
1848+04 & 0.285 &18.6 & 19.3\\
1859+01 & 0.288 & 4.6 & 4.8\\
1904+06 & 0.267 & 9.7 & 10.0\\
1918+26 & 0.785 & 4.1 & 4.2\\
1944+22 & 1.334 & 6.5 & 6.7\\
2027+37 & 1.217 & 4.8 & 5.0\\
2034+19 & 2.075 & 3.6 & 3.7\\
2035+36 & 0.618 & 3.6 & 3.7\\ \hline\hline
\multicolumn{4}{|c|}{WEISBERG 2004 -- 0.430 GHz}\\ \hline
Name B  & $P$ & $W_{50}$ (0.43 GHz)& $W_{50}$ (1 GHz)\\
&  [s]& [$^{\circ}$] & [$^{\circ}$]  \\ \hline
0045+33 & 1.218 &  5.6 & 3.2\\
1915+22 & 0.426 & 15.6 &14.3\\
1922+20 & 0.238 & 17.0 &15.6\\
1929+15 & 0.314 & 10.8 & 9.9\\ \hline\hline
\multicolumn{4}{|c|}{PARKES I -- 1.374 GHz}\\ \hline
Name J & $P$ &$W_{50}$ (1.374 GHz) & $W_{50}$ (1 GHz)\\
& [s] & [$^{\circ}$]  & [$^{\circ}$]  \\ \hline
0957-5432&0.2035&6.4 & 6.6\\
1049-5833&2.2022&5.4 & 5.6 \\
1123-6102&0.6402&5.6 & 5.8\\
1144-6217&0.8506&4.2 & 4.3\\
1252-6314&0.8233&8.8 & 9.1\\
1348-6307&0.9278&11.0& 11.4\\
1349-6130&0.2594&22.2& 22.9\\ \hline

\newpage

\multicolumn {4}{l}{Tab. \ref{tab.c2} cont.}\\
\hline
\multicolumn{4}{|c|}{PARKES IV -- 1.374 GHz}\\ \hline
Name J &  $P$ & $W_{50}$ (1.374 GHz)& $W_{50}$ (1 GHz)\\
& [s] & [$^{\circ}$]  & [$^{\circ}$]  \\ \hline
0942-5552& 0.6644 &9.    &9.3 \\
1121-5444&0.5358  & 14.1 &14.6 \\
1157-6224&0.4005  & 11.7 &12.1 \\
1225-6408&0.4196  & 6.9  &7.1 \\
1326-5859&0.4780  & 5.7  &5.9 \\
1534-5408&0.2897  & 7.6  &7.8 \\
1703-4851&1.3964  & 3.4  &3.5 \\
1749-3002&0.6099  & 21.  &21.7 \\
1801-2920&1.0818  & 8.   &8.3 \\
1807-0847&0.1637  & 9.   &9.3 \\
1849-0636&1.4514  & 7.4  &7.6 \\
1857+0057&0.5369  & 4.9  &5.1 \\
1909+0007&1.0169  & 2.8  &2.9 \\
1909+0254&0.9899  & 4.0  &4.1 \\
1910-0309&0.5046  & 5.4  &5.6 \\  \hline\hline
\multicolumn{4}{|c|}{PARKES VI -- 1.374 GHz}\\ \hline
Name J & $P$ &$W_{50}$ (1.374 GHz)& $W_{50}$ (1 GHz)\\
& [s] &[$^{\circ}$]   & [$^{\circ}$]  \\ \hline
1717-3953& 1.0855&31. & 32.0  \\
1728-4028& 0.8663&5.6 &  5.8  \\
1822-0848& 0.8348&10. &  10.3 \\
1827-0750& 0.2705&12.5&  12.9 \\  \hline
\end{longtable}

\twocolumn

\label{lastpage}

\end{document}